\documentstyle[12pt]{article}
\def\PRL #1 #2 #3{{\sl Phys. Rev. Lett.} {\bf#1} (#2) #3}
\def\NPB #1 #2 #3{{\sl Nucl. Phys.} {\bf B#1} (#2) #3}
\def\NPBFS #1 #2 #3 #4{{\sl Nucl. Phys.} {\bf B#2} [FS#1] (#3) #4}
\def\CMP #1 #2 #3{{\sl Commun. Math. Phys.} {\bf #1} (#2) #3}
\def\PRD #1 #2 #3{{\sl Phys. Rev.} {\bf D#1} (#2) #3}
\def\PLA #1 #2 #3{{\sl Phys. Lett.} {\bf #1A} (#2) #3}
\def\PLB #1 #2 #3{{\sl Phys. Lett.} {\bf #1B} (#2) #3}
\def\JMP #1 #2 #3{{\sl J. Math. Phys.} {\bf #1} (#2) #3}
\def\PTP #1 #2 #3{{\sl Prog. Theor. Phys.} {\bf #1} (#2) #3}
\def\SPTP #1 #2 #3{{\sl Suppl. Prog. Theor. Phys.} {\bf #1} (#2) #3}
\def\AoP #1 #2 #3{{\sl Ann. of Phys.} {\bf #1} (#2) #3}
\def\PNAS #1 #2 #3{{\sl Proc. Natl. Acad. Sci. USA} {\bf #1} (#2) #3}
\def\RMP #1 #2 #3{{\sl Rev. Mod. Phys.} {\bf #1} (#2) #3}
\def\PR #1 #2 #3{{\sl Phys. Reports} {\bf #1} (#2) #3}
\def\AoM #1 #2 #3{{\sl Ann. of Math.} {\bf #1} (#2) #3}
\def\UMN #1 #2 #3{{\sl Usp. Mat. Nauk} {\bf #1} (#2) #3}
\def\FAP #1 #2 #3{{\sl Funkt. Anal. Prilozheniya} {\bf #1} (#2) #3}
\def\FAaIA #1 #2 #3{{\sl Functional Analysis and Its Application} {\bf
#1} (#2) #3}
\def\BAMS #1 #2 #3{{\sl Bull. Am. Math. Soc.} {\bf #1} (#2)
#3} \def\TAMS #1 #2 #3{{\sl Trans. Am. Math. Soc.} {\bf #1} (#2) #3}
\def\InvM #1 #2 #3{{\sl Invent. Math.} {\bf #1} (#2) #3}
\def\LMP #1 #2 #3{{\sl Letters in Math. Phys.} {\bf #1} (#2) #3}
\def\IJMPA #1 #2 #3{{\sl Int. J. Mod. Phys.} {\bf A#1} (#2) #3}
\def\AdM #1 #2 #3{{\sl Advances in Math.} {\bf #1} (#2) #3}
\def\RMaP #1 #2 #3{{\sl Reports on Math. Phys.} {\bf #1} (#2) #3}
\def\IJM #1 #2 #3{{\sl Ill. J. Math.} {\bf #1} (#2) #3}
\def\APP #1 #2 #3{{\sl Acta Phys. Polon.} {\bf #1} (#2) #3}
\def\TMP #1 #2 #3{{\sl Theor. Mat. Phys.} {\bf #1} (#2) #3}
\def\JPA #1 #2 #3{{\sl J. Physics} {\bf A#1} (#2) #3}
\def\JSM #1 #2 #3{{\sl J. Soviet Math.} {\bf #1} (#2) #3}
\def\MPLA #1 #2 #3{{\sl Mod. Phys. Lett.} {\bf A#1} (#2) #3}
\def\JETP #1 #2 #3{{\sl Sov. Phys. JETP} {\bf #1} (#2) #3}
\def\JETPL #1 #2 #3{{\sl  Sov. Phys. JETP Lett.} {\bf #1} (#2) #3}
\def\PHSA #1 #2 #3{{\sl Physica} {\bf A#1} (#2) #3}
\def\CQG #1 #2 #3{{\sl Class. Quantum Grav.} {\bf #1} (#2) #3}
\def\SJNP #1 #2 #3{{\sl Sov. J. Nucl. Phys. (Yadern.Fiz.)} {\bf #1} (#2) #3}
\def\a{\alpha}\def\b{\beta}\def\d{\delta}

\def\k{\kappa}
\def\G{\Gamma}
\def\G{\Gamma}
\def\u{\underline}
\def\be{\begin{equation}}\def\ee{\end{equation}}

\newcommand{\bea}{\begin{eqnarray}}
\newcommand{\eea}{\end{eqnarray}}
\newcommand{\nn}{\nonumber\\}\newcommand{\p}[1]{(\ref{#1})}
\begin{document}
\renewcommand{\thefootnote}{\arabic{footnote}}
\begin{flushright}
{HUB-EP-98/37\\
DFPD 98/TH/30\\
hep-th/9807050}
\end{flushright}
\begin{center}

\bigskip
{\large\bf On Some Features of the M--5--Brane
\footnote{Invited talk given at the Trieste Conference on Superfivebranes\\
and Physics in 5+1 Dimensions (April 1--3, 1998), and at the 2nd INTAS 
Meeting ``Fundamental Problems in Classical, Quantum and String Gravity"
(Paris, 1--3 May, 1998)}
}

\vspace{1cm}
{Dmitri Sorokin}\footnote{Alexander von Humboldt Fellow\\
\,\, \, On leave from Kharkov Institute of Physics and
Technology, Kharkov, 310108, Ukraine}

\bigskip
Humboldt-Universit\"at zu Berlin,
Institut f\"ur Physik\\
Invalidenstrasse 110, D-10115 Berlin,\\
Bundesrepublik Deutschland\\
e-mail: sorokin@qft2.physik.hu-berlin.de

\bigskip
and

\bigskip
Universit\`a Degli Studi Di Padova\\
Dipartimento Di Fisica ``Galileo Galilei''\\
ed INFN, Sezione Di Padova\\
Via F. Marzolo, 8, 35131 Padova, Italia

\vspace{1cm}
{\bf Abstract}

\bigskip
We review the structure and symmetry properties of the worldvolume action
for the M--theory 5--brane and of its equations of motion.
\end{center}

\newpage
The super--five--brane is one of fundamental objects of M--theory, and
its existence reflects or causes important duality chains connecting $D=11$
supergravity with $D=10$ string theories, and string theories among 
themselves. A complicated dynamics of 5-branes also gives rise to a new
important class of $d=6$ superconformal field theories, which, in their 
own turn, are related to four--dimensional quantum $N=2$ super--Yang--Mills
theories.

Thus, understanding various features of classical dynamics of the 5--brane
should be useful for further developments in this field of research.

Some of these properties, especially the structure of 5--brane symmetries
have found to be rather peculiar and unexpected from the point of view of
previous superbrane experience. And in this contribution I would like to
discuss what we have learned about properties of the 5--brane of
M--theory by studying its worldvolume action.

The points to be considered are:\\
i) the structure and symmetries of the M--5--brane action;\\
ii) M--theory supertranslation algebra as an algebra of Noether charges
of the 5--brane;\\
iii) 5--brane equations of motion an their relation to geometrical
conditions of 
superembedding 5--brane worldvolume into $D=11$ target superspace.\\

A 5--brane was first observed as a solitonic solution of $D=11$ 
supergravity equations of motion \cite{gu}. It was realized that this
soliton preserves half the supersymmetry of a $D=11$ vacuum (i.e. 16 
supercharges of 32). Hence, an effective theory describing small fluctuations of 
the 5--brane should be a $d=6$ worldvolume theory with 16 linearly realized 
(chiral) supersymmetries. A linear on--shell $d=6$  supermultiplet of this 
supersymmetry consists of 5 scalars, a self--dual (or chiral) two--form field,
which carries 3 physical degrees of freedom and 8 fermions. The 5 scalars and
8 fermions are associated with fluctuations of the 5--brane in target $N=1$
$D=11$ superspace along directions transversal to the five--brane worldvolume,
while the self--dual field intrinsically propagates in the worldvolume.

The presence of this chiral field caused a main problem for the construction of 
the 5--brane action. This problem is generic for the Lagrangian description
of all self--dual fields and is twofold:

i) one should  construct an action which produces the self--duality condition
as an equation of motion, and

ii) keep space--time covariance of the construction manifest. 

The latter is desirable, since, as usual, space--time covariance substantially 
simplifies coupling the fields to gravity and supergravity.

Before presenting the complete 5--brane action consider how both these 
problems can be solved in the case of a free two--form field $B_{mn}(y)$
$(m,n = 0,1,...,5)$ in $d=6$, whose three--form field strength $H^{(3)}=dB^{(2)}$ 
is self--dual on the mass shell \cite{pst}
\begin{equation}\label{1}
H^{m_1m_2m_3}=H^{*m_1m_2m_3}=
{1\over 6}\varepsilon^{m_1m_2m_3m_4m_5m_6}H_{m_4m_5m_6}
\end{equation}
A covariant action \cite{pst}, which yields \p{1} as a consequence of 
the $B^{(2)}$
equation of motion, can be written in two equivalent forms
\begin{equation}\label{2}
S=\int d^6 y [-{1\over {4!}}H^{mnp}H_{mnp} + {1\over 8} 
(H^{*mn}-H^{mn})(H^*_{mn}-H_{mn})],
\end{equation}
\begin{equation}\label{3}
S=\int d^6 y (-{1\over 4}H^{*mn}H^*_{mn} + {1\over 4} H^{*mn}H_{mn}),
\end{equation}
where  by definition $H_{mn}\equiv H_{mnp}v^p$, $H^*_{mn}\equiv H^*_{mnp}v^p$, 
and the  vector $v_p={{\partial_pa}\over{\sqrt{-(\partial a)^2}}}$ 
is a normalized gradient $v^pv_p=-1$ 
of an auxiliary scalar field $a(y)$ which is in charge of the manifest 
Lorentz covariance of the actions.

The first form \p{2} of the action clearly exhibits the difference of the 
self--dual action from the action of a nonchiral two--form field. This 
difference is in the presence of an additional term in \p{2} 
quadratic in the anti--self--dual part of $H^{(3)}$.

The second form of the action \p{3} turns out to be the most suitable for the
generalization to a complete non--linear 5--brane action, as we shall see below.

In addition to ordinary gauge symmetry $\delta B^{(2)}=d\varphi^{(1)}(y)$ the 
action \p{2}, \p{3} is invariant under two more symmetries with a one--form and
scalar parameter, respectively:
\begin{equation}\label{4}
\delta B^{(2)}=da\wedge\phi^{(1)}(y), \qquad \delta a(y)=0;
\end{equation}
\begin{equation}\label{5}
\delta a=\phi(y), \qquad \delta B_{mn}=
{{\phi(y)}\over{\sqrt{-(\partial a)^2}}}(H^*_{mn}-H_{mn}).
\end{equation}
The first of these symmetries is responsible for the fact that the second--order
differential equation
\begin{equation}\label{5.1}
{{\delta S}\over{\d B_{mn}}}=\epsilon^{mnpqrs}\partial_p(v_q H^*_{rs})=0
\end{equation}
reduces to the first--order self--duality condition \p{1} (see \cite{pst} for 
details).
One should anticipate the presence of this symmetry in the self--dual action 
since the self--dual (chiral) field carries twice less physical degrees of 
freedom than the non--chiral one, and, hence, there should be more local 
symmetries to eliminate redundant degrees of freedom contained in $B_{mn}$.

The third local symmetry \p{5} reflects an auxiliary nature of the scalar field
$a(y)$ and can be used to gauge fix $\partial_m a(y)$ to be either time--like
or space--like constant vector at the expense of  
manifest space--time covariance. 
As we shall see, for different applications it 
is convenient to make one or another choice of the constant vector.

When coupled to $d=6$ gravity induced by embedding of a six manifold to 
$D=11$ 
space--time, the action \p{2}, \p{3} can be regarded as a quadratic 
approximation of a highly non--linear 5--brane worldvolume action. 
The assertion 
that the complete action should be nonlinear and describe a 
Dirac--Born--Infeld--like (DBI) self--interaction of the chiral field 
$B^{(2)}$
is based on an observation that double dimensional reduction of a $D=11$ 
5--brane down to $D=10$ (i.e. when one of the 5--brane coordinates is wrapped 
around the compactified 11-th dimension) must result in a Dirichlet--4--brane
whose worldvolume action has been known to be of a DBI--type.

A break through in constructing the complete 5--brane action was made when 
a DBI generalization of the action for a chiral two--form 
field was constructed in
a non--covariant form in \cite{ps} and extended to a covariant action for a
bosonic 5--bane in \cite{pst1}. The prescription how to do this now looks
surprisingly straightforward. One should simply replace the first quadratic
term in \p{3} with a DBI--like term and couple $B^{(2)}$ to a $d=6$ metric
$g_{mn}(y)=\partial_mX^{\u p}(y)E^{\u a}_{\u p}E_{\u{a}\u{q}}
\partial_n X^{\u q}(y)$
induced by embedding into $D=11$ target space. 
$E^{\u a}=dX^{\u m}E^{\u a}_{\u 
m}(X)$ is a $D=11$ vielbein one--form determining a basis in $D=11$ tangent
space. Underlined indices from the beginning of the alphabet $(\u a, \u b 
=0,1,...,10)$  are $SO(1,10)$ Lorentz indices and 
$(\u m, \u n ...=0,1,...,10)$
are indices of $D=11$ curved coordinates.

As a result, we get the 5--brane action in the following form
\begin{equation}\label{6}
S=\int d^6 y (-\sqrt{-\det(g_{mn}+{H}^*_{mn})}
+ {\sqrt{-g} \over 4}
H^{*mn} H_{mn}),
\end{equation}
where now $H^{*m_1m_2m_3}={1\over{6\sqrt{-g}}}\epsilon^{m_1m_2m_3m_4m_5m_6}
H_{m_4m_5m_6}$.

The action \p{6} is $d=6$ general--coordinate invariant, and possesses local 
symmetries analogous to \p{4} and \p{5} (which is important for consistency).
Namely, the symmetry \p{4} is the same, while the variation of $B^{(2)}$ in
\p{5} gets modified and takes the form
\begin{equation}\label{7}
\delta a(y)=\phi(y), \qquad \delta B_{mn}={{\phi(y)}\over{\sqrt{(\partial 
a)^2}}}({{2\delta L_{DBI}}\over{\delta H^{*mn}}}-H_{mn}),
\end{equation}
$$
L_{DBI}=\sqrt{\det(\d_{m}^{n}+{H}^{*~n}_{m})}.
$$
It is useful to note that the form of the $B^{(2)}$--variation prompts the form
of the generalized self--duality condition on $H^{(3)}$, which now reads that
on the mass shell
\begin{equation}\label{8}
H_{mn}\equiv H_{mnp}v^p={{2\delta L_{DBI}}\over{\delta H^{*mn}}}=
{{H^*_{mn}-{1\over 2}tr(H^*)^2H^*_{mn}+(H^*)^3_{mn}}\over{L_{DBI}}}.
\end{equation}
Eq. \p{8} reduces to eq. \p{1} in a linear approximation.

Note that \p{8} explicitly contains the auxiliary vector
$v_p={{\partial_pa}\over{\sqrt{-(\partial a)^2}}}$, however, as was shown in 
\cite{hsw1}, the generalized self--duality condition can also be reformulated
in terms of $H^{(3)}$ and ${}^*H^{(3)}$ only (see eq. (24) of ref. \cite{hsw1}
for details).

Let us now couple the 5--brane action \p{6} to a three form gauge field 
$A^{(3)}(X)$ of $D=11$ supergravity. We should keep in mind that such coupling
must not spoil local symmetries \p{4} and \p{7} which guarantee the 
self--duality properties of the 5--brane. 

By analogy with a D--brane coupling we may assume that the worldvolume pullback
$A_{mnp}(X(y))=A_{\u{mnp}}(X(y))\partial_m X^{\u m}\partial_n X^{\u n}
\partial_p X^{\u p}$ of the gauge field enters the 5--brane action through the 
modified field strength
\begin{equation}\label{9}
{\hat H}^{(3)}=H^{(3)}-A^{(3)}=dB^{(2)}-A^{(3)}.
\end{equation}
We should require ${\hat H}^{(3)}$ to be invariant under $D=11$ gauge 
symmetry
$\delta A^{(3)}=d\phi^{(2)}$ pulled back into the 5--brane worldvolume.
This takes place if $B^{(2)}$ is shifted by this symmetry
\begin{equation}\label{10}
\delta B^{(2)}=\varphi^{(2)}(X(y)), \qquad \delta A^{(3)}=d\varphi^{(2)}(X(y)).
\end{equation}
If we now substitute ${\hat H}^{(3)}$ instead of $H^{(3)}$ into eq. \p{6}
and check whether such an action is invariant under \p{4} and \p{7}, we can 
easely 
find that the corresponding
variation does not vanish, and the nonvanishing term is
\begin{equation}\label{11}
\d S=-\int_{{\cal M}_6}{1\over 2}\delta B^{(2)}\wedge dA^{(3)}.
\end{equation}
It can be canceld by adding to the action \p{6} the term
\begin{equation}\label{12}
S_1=-\int_{{\cal M}_6}{1\over 2}d B^{(2)}\wedge A^{(3)}.
\end{equation}
But this is not the end of the story since \p{12} is not invariant under
the gauge transformations \p{10}.
To compensate this nonivariance the 5--brane should minimally couple
to a $D=11$ six--form field $A^{(6)}(X)$ dual to $A^{(3)}$, the variation
of $A^{(6)}$ under the $D=11$ gauge transformations being 
$$
\delta A^{(6)}=d\varphi^{(5)}+\varphi^{(2)}\wedge dA^{(3)}.
$$
As a result we recover the 5--brane Wess--Zumino term of Aharony \cite{ah}
\begin{equation}\label{13}
S_{WZ}= -\int_{{\cal M}_6}{1\over 2}(A^{(6)}+d B^{(2)}\wedge A^{(3)}).
\end{equation}
The fields $A^{(6)}$ and $A^{(3)}$ are dual to each other in the sense that
their field strengths are related through the following condition
\begin{equation}\label{du}
F^{(7)}\equiv dA^{(6)}-A^{(3)}\wedge dA^{(3)}={~}^*dA^{(3)}\equiv {~}^*F^{(4)},
\end{equation}
where ${~}^*$ denotes the $D=11$ Hodge operation.
The relation \p{du}
has been determined such that it
implies the equations of motion and Bianchi identities for the field
$A^{(3)}$ which follow from the $D=11$ supergravity action \cite{cjs}.

Since the field $A^{(6)}$ is not present in the standard formulation of
$D=11$ supergravity, one may wonder if the $A^{(6)}$ term in the Wess--Zumino 
action can be replaced with another term, which would depend on the field
$A^{(3)}$. The only known form \cite{alwis,bbs} of such a term is following
\begin{equation}\label{131}
S_{WZ}= -\int_{{\cal M}_7}{1\over 2}A^{(3)}\wedge dA^{(3)}-
\int_{{\cal M}_6}{1\over 2}d B^{(2)}\wedge A^{(3)},
\end{equation}
where the first term is an integral over a 7-manifold whose boundary is
the 5--brane worldvolume ${\cal M}_6$. This nonlocal coupling of the 
5--brane to the three--form field is a manifestation of a magnetic nature
of the 5--brane, and is similar to nonlocality of monopole coupling
in $D=4$ electrodynamics. ${\cal M}_7$ is associated with a Dirac 6--brane
stemmed from the 5--brane.
Here we shall not elaborate this version in more detail, and will consider the
formulation where the dual six--form field $A^{(6)}$ is explicitly present.
Note that there exists a corresponding version of $D=11$ supergravity \cite{bbs} 
where both the $A^{(3)}$  and $A^{(6)}$  field enter the action in a 
duality--symmetric form.

We have thus constructed the complete worldvolume action for the five--brane 
coupled to a bosonic background of $D=11$ supergravity
\begin{equation}\label{14}
S=\int d^6 y (-\sqrt{-\det(g_{mn}+{\hat H}^*_{mn})}
+ {\sqrt{-g} \over 4}
\hat H^{*mn} \hat H_{mn})-
\int_{{\cal M}_6}{1\over 2}(A^{(6)}+d B^{(2)}\wedge A^{(3)}).
\end{equation}

A novel feature which we have observed is that the local worldvolume 
symmetries of the 5--brane responsible for its self--duality properties require
the presence in the action 
of the Wess--Zumino term for the $D=11$ gauge field coupling to 
be consistent. In the case of all other superbranes that has been a prerogative
of $\kappa$--symmetry of the full supersymmetric action.

In the present case, when all terms of the 5--brane action have been completely 
fixed already at the bosonic level it is straightforward to generalize it
to a supersymmetric action describing a five--brane propagating in a $D=11$ 
supergravity background \cite{pst2,s1}. 
For this we should only replace $D=11$ background
fields $E^{\u a}(X)$, $A^{(3)}(X)$  and $A^{(6)}(X)$ with corresponding 
superforms in target superspace parametrized by supercoordinates 
$Z^{M}=(X^{\u m},\Theta^\alpha)$, where $\Theta^\alpha$ is a 32--component
Majorana spinor. Thus, the target--space supersymmetric action for the
5--brane is the same as eq. \p{14}, but where now
$$
g_{mn}(y)=\partial_mZ^ME_M^{\u a}E_{{\u a}N}\partial_nZ^N, \qquad 
E^A_m=\partial_mZ^ME_M^A,
$$
$$
A^{(3)}=A_{M_1M_2M_3}dZ^{M_3}(y)dZ^{M_2}(y)dZ^{M_1}(y),
$$
\begin{equation}\label{15}
A^{(6)}=A_{M_1...M_6}dZ^{M_6}(y)...dZ^{M_1}(y).
\end{equation}
The differential superforms in \p{15} are Grassmann supersymmetric
(i.e. antisymmetric with respect to bosonic coordinates and symmetric with 
respect to fermionic coordinates).

An important and tricky point has been to check that the supersymmetric action
\p{14} possesses also a worldvolume fermionic $\kappa$--symmetry, and hence 
describes BPS 5--brane configurations preserving half the 
$D=11$ supersymmetry.
The experience of studying other super--p--branes prompts us that the
$\kappa$--symmetry variation of worldvolume fields should be of the following
form
$$
i_\kappa E^{\alpha}\equiv \delta_\kappa Z^{M}E_M^\alpha=
(1+\bar\Gamma)_{~\beta}^{\alpha}\kappa^\beta, \qquad i_\kappa E^{\u a}\equiv 
\delta_\kappa Z^{M}E_M^{\u a}=0, 
$$
\begin{equation}\label{16}
\delta g_{mn}=-4iE_{\{m}^\alpha (\Gamma_{n\}})_{\alpha\beta}i_\kappa E^\beta,
\qquad \delta \hat H^{(3)}=-i_\kappa dA^{(3)},\qquad \delta_\kappa a(y)=0,
\end{equation}
where $\kappa^\alpha(y)$ is a fermionic parameter, 
$i_\kappa$ denotes the contraction of form components with 
$\delta_\kappa Z^{M}$,
$\Gamma_{m_1...m_p}$ denotes 
an antisymmetric product of $p$
$D=11$ gamma--matrices pulled back into the d=6 worldvolume, i.e.
$\Gamma_m=\Gamma_{\u a}E^{\u a}_m(Z(y))$, and 
$1+\bar\Gamma(y)$ is a projector matrix with $\bar\Gamma^2=1$.
Therefore, only 16 components of the 32--component parameter $\kappa^\alpha(y)$
effectively participate in the transformations \p{16}, and this reflects the 
fact that half the 32 supersymmetries of a $D=11$ vacuum are broken because
of the presence of the 5--brane.

It turns out that $\bar\Gamma$ is not uniquely defined. So the proof of 
$\kappa$--invariance becomes a relatively simple technical problem when
an appropriate form of the projector $1+ \bar\Gamma(y)$ is found.
The following $\bar\Gamma$ appears to be the most suitable one
\bea
\bar\G =&& - {1 \over \sqrt{-\det(g_{mn}+\hat H^*_{mn})}}
\bigg[(v_m \G^m)\G_nt^n +
{\sqrt{-g}\over 2}\,
(v_m \G^m)\G^{np} \hat H^*_{np} \nn
&& +\qquad {1\over 5!}\,
(v_m  \G^m) \G_{i_1\dots i_5}\,
{\varepsilon}^{m_1\dots m_5 n} v_n \bigg]\, ,
\label{17}
\eea
where (note that $t^m v_m \equiv 0$)
\be
\label{171}
t^m = {1\over 8}\,
{\varepsilon}^{mn_1n_2p_1p_2 q}
{\hat H}^*_{n_1n_2}{\hat H}^*_{p_3p_4} v_q 
\ee

As common to all super--p--branes the $\kappa$--symmetry of the brane action
requires the background superfields to satisfy constraints.
In our case the $\kappa$--symmetry 
is compatible with $D=11$ supergravity constraints (see, for example,
\cite{cl}) which put the background on the mass shell
$$
T^{\u a}_{\alpha\beta}=-i \Gamma^{\u a}_{\alpha\beta}~~~({\rm torsion
~components}),
$$
$$
dA^{(3)} ={i\over 2}E^{\u a}  E^{\u b} E^\a  E^\b (\Gamma_{\u{ab}})_{\a\b}
+{1\over {4!}}E^{\u a}  E^{\u b} E^{\u c}E^{\u d} F^{(4)}_{\u{dcba}}
$$
\be\label{18}
dA^{(6)}-A^{(3)}dA^{(3)} 
={{2i}\over {5!}}E^{{\u a}_1}\dots E^{{\u a}_5}  E^\a  E^\b
(\Gamma_{{\u a}_1\dots {\u a}_5})_{\a\b}
 + {1\over {7!}}E^{{\u a}_1} \dots E^{{\u a}_7} F^{(7)}_{{\u a}_7\dots{\u 
a}_1}\, ,
\ee
where the purely vector vielbein components of the field strengths
of the superfields $A^{(3)}$ and $A^{(6)}$ are Hodge dual to each other
$F^{(7)}={~}^*F^{(4)}$ as in the bosonic case \p{du}.

When target superspace is flat, then, in agreement with the constraints \p{18}
\begin{equation}\label{181}
E^{\u a}=dX^{\u a}+id\bar\Theta\Gamma^{\u a}\Theta,\qquad E^\alpha=d\Theta^\alpha,
\qquad F_{{\u m}_1...{\u m}_7}=F_{{\u m}_1...{\u m}_4}=0.
\end{equation}
The explicit form of $A^{(3)}$ and $A^{(6)}$ in flat $D=11$ superspace is rather
complicated \cite{t,ka}. Fortunately it is not required for further analysis of 
the 5--brane action.

Having constructed the action one should check whether it really describes the 
5--brane which we are interested in, i.e. the 5--brane of M--theory.
For instance, whether the action \p{14} reduces to an action for a Dirichlet
4--brane of type IIA superstring theory upon the double dimensional reduction, 
when one of the 5--brane coordinates is wrapped around the compactified 
direction of $D=11$ space--time. This has been checked in several papers
\cite{ps,pst1,s1,s2}. It was shown that upon the dimensional reduction one gets
a dual formulation of the D4--brane, where instead of a vector gauge field
$A^{(1)}_i$ (i=0,1,...,4) the D4--brane carries a two--form field $B_{ij}$.
On the mass shell the field strength of $ A^{(1)}$ and $B^{(2)}$ are related
by $d=5$ Hodge duality $F_{ij}=H^*_{ij}={1\over 
6}\varepsilon_{ijk_1k_2k_3}H^{k_1k_2k_3}$. The standard form \cite{dbrane}
of the D4--brane action is recovered upon an (anti)dualization procedure, 
discussed, for instance, in \cite{ts}.

A duality relation of a 5-brane compactified on $K3$ with a heterotic string 
compactified on a torus 
was considered in \cite{cher}.

When performing the dimensional reduction of the 5--brane action, it is 
convenient to make use of the local symmetry \p{7} associated with a local
shift of the auxiliary field $a(y)$ and identify $a(y)$ with compactified 
coordinates, for instance, $X^{10}=y^5=a$. In this gauge one straightforwardly 
observes that $a(y)$ disappears from the covariant ten--dimensional D4--brane
action.

Another interesting question is what kind of $D=11$ superalgebra is generated
by 5--brane Noether currents associated with supertranslations in flat $D=11$
superspace. This analysis was carried out in \cite{st} and it was shown that the
5--brane Noether supercharges $Q_\alpha$, $P_{\u m}$ form the most general 
supertranslation algebra in eleven dimensions called the M--theory superalgebra
\be
\label{19}
\{ Q_{\a}, Q_{\b} \} =(\G^0\G^{\u m})_{\a \b} P_{\u m} +
{1\over 2}(\G^0\G^{\u{mn}})_{\a \b}Z_{\u{mn}}+ {1\over 5!}
(\G^0\G ^{{\u m}_1\dots {\u m}_5})_{\a \b}Y_{{\u m}_1\dots {\u m}_5},
\ee
where $Z_{\u{mn}}$ and $Y_{{\u m}_1\dots {\u m}_5}$ are a two--form and a 
five--form central charge. (We use the Majorana representation of the 
$\gamma$--matrices where $\G^0$ plays the role of the charge conjugation
matrix).

A usual assertion concerning the nature of these central
charges is that the two--form $Z_{\u{mn}}$ is associated solely with an 
``electric" membrane minimally coupled to the three--form potential $A^{(3)}$
via the membrane Wess--Zumino term $\int_{{\cal M}_3}A^{(3)}$. 
And the five--form charge
is associated with a magnetically dual 5--brane minimally coupled to the 6--form
potential $A^{(6)}$. However, the structure of the 5--brane action tells us
that the 5--brane couples to both $A^{(3)}$ and $A^{(6)}$ and, hence, is
a dyonic object, which carries both the two--form and the five--form charge.
And these charges appear on the right hand side of the superalgebra of the
5--brane Noether charges. Let us consider this in more detail.

To get the form of the supersymmetry generators $Q_\alpha$ as Noether charges,
one applies the standard Noether prescription, which in our case consists in 
performing the variation of the action with respect to supersymmetry 
transformations in flat target superspace and taking into account that the 
Wess--Zumino term is invariant only up to a total derivative
$$
\delta S_{WZ}=\int_{{\cal M}_6}id\bar\epsilon
(2\Delta^5-\Delta^2\wedge {\hat H}^{(3)}).
$$
The supersymmetry
variations of fields have the following form:
$$
\delta\Theta=\epsilon, \qquad \delta X^{\u m}=i\bar\epsilon\Gamma^{\u m}\Theta,
\qquad \delta B^{(2)}=i\bar\epsilon\Delta^2,
$$
\begin{equation}\label{20}
\delta A^{(3)}=id(\bar\epsilon\Delta^2), \qquad 
\delta A^{(6)}=2id(\bar\epsilon\Delta^5)+id(\bar\epsilon\Delta^2)\wedge dA^{(3)},
\end{equation}
where $\Delta^p$ (p=1,5) are spinor valued p-forms whose relevant leading terms 
in $\Theta$--expansion are
\begin{equation}\label{21}
\Delta^p={1\over{p!}}\Gamma_{{\u m}_1...{\u m}_p}
\Theta dX^{{\u m}_p}...dX^{{\u m}_1}
~+~...
\end{equation}
Note that the worldvolume gauge field $B^{(2)}$ transforms nontrivially under 
$D=11$ supersymmetry, its variation being proportional to the worldvolume 
pullback of $\Delta^2$. This is required for the field strength $\hat H^{(3)}=
dB^{(2)}-A^{(3)}$ to be superinvariant.
Therefore, the supercharge generator must act on the field $B^{(2)}$ 
nontrivially. This is reflected in the form of the Noether supercharge
\bea
Q_\alpha&=& i\int\! d^5\sigma\,\big[(\pi+i\bar\Theta\Gamma^{\u m}{\cal 
P}_{\u m})_\alpha
+i({\cal P}^{i_1i_2} + {1\over 4}
{H}^{*0i_1i_2})(\Delta^2_{{i_1i_2}})_\alpha
-i\varepsilon^{i_1...i_5}
(\Delta^5_{i_1...i_5})_\alpha\big]\, .
\label{22}
\eea
where the integral is taken over a 5--brane surface $(i=1,...,5)$, and
$\pi_\alpha$, ${\cal P}_{\u m}$ and ${\cal P}^{i_1i_2}$ are canonical conjugate 
momentum densities of $\Theta^\alpha$, $X^{\u m}$ and $B_{ij}$, respectively.
They are defined as corresponding variations of the Lagrangian 
${{\delta L}\over{\delta(\partial_0\phi)}}$ with respect to field velocities.

We see that the form of the supercharge $Q_\alpha$ is different from the one 
which we get accustomed to in the superfield theory or in the case of ordinary
super--p--branes. It contains the standard term with the $\Theta$-- and 
$X$--momentum, and it also contains the canonical momentum density of the gauge
field $B^{(2)}$, and two other terms. These last two terms appear in $Q_\alpha$
because of the noninvariance of the Wess--Zumino part 
of the 5--brane action \p{14}.
Note that Noether supercharges of the type II $D=10$ Dirichlet branes should 
have the analogous structure and contain a contribution of the worldvolume 
momentum of the vector gauge field, since it also varies under supersymmetry 
transformations \cite{hammer}.

Taking the anticommutator of two 5--brane supercharges we obtain that it 
reproduces
the r.h.s. of the M--theory superalgebra \p{19} with $P_{\u m}$, $Z_{\u{mn}}$ 
and $Y^{{\u m}_1\dots {\u m}_5}$ having the following form
\begin{equation}\label{23}
P_{\u m}=\int d^5\sigma {\cal P}_{\u m},
\end{equation}
\begin{equation}\label{24}
Z^{\u{mn}}=-2\int_{{\cal M}_5} dX^{\u m}\wedge dX^{{\u n}}\wedge({\cal 
P}^{*(3)}+
{1\over 4}dB^{(2)}) =- \int_{{\cal M}_5} dX^{\u m}\wedge dX^{{\u 
n}}\wedge dB^{(2)},
\end{equation}
\begin{equation}\label{25}
Y^{{\u m}_1\dots {\u m}_5}=\int_{{\cal M}_5} dX^{{\u m}_1}\dots dX^{{\u m}_5}.
\end{equation}
The last expression in \p{24} for the two--form central charge is obtained by an 
explicit computation of the dual momentum density of the gauge field $B^{(2)}$
\begin{equation}\label{26}
{\cal P}^*_{i_1i_2i_3}\equiv{1\over 2}\varepsilon_{i_1i_2i_3i_4i_5}
{{\delta L}\over{\delta \partial_0 B^{i_4i_5}}}={1\over 4}\partial_{[i_1}
B_{i_2i_3]}.
\end{equation}
The easiest way to get this expression is to choose the temporal gauge 
$a(y)=y^0=\tau$ for the auxiliary field $a(y)$ in the five--brane action
(where $\tau$ is a proper worldvolume time). Then the Dirac--Born--Infeld part
of the action does not contribute to the definition of ${\cal P}^{ij}$.
We can also notice that eq. \p{26} does not contain time derivatives and hence
is a Hamiltonian constraint which reflects the self--duality of $B^{(2)}$.
A complete Hamiltonian for the M--5--brane was recently constructed in 
\cite{bst}.

Note that only half value of $Z^{(2)}$ is due to the contribution of
the Wess--Zumino term, and another half comes from the $\hat H^{(3)}$--quadratic
term of the 5--brane action \p{14}. In the case of all other branes the 
central charges in superalgebras are associated solely with corresponding
Wess--Zumino terms.

 From the expressions \p{24} and \p{25} for $Z$ and $Y$ we see that 
(in accordance with a general observation \cite{t}) they are 
topological charges which are non--zero only for topologically nontrivial 
configurations of the 5--brane worldvolume and the self--dual gauge field.
For instance, $H^{(3)}=dB^{(2)}$ should be closed but not exact. 
A simple example is an 
infinite planar five--brane in flat $D=11$ space--time with $dB^{(2)}$
being a constant three--form.

Another observation one can make is that the form of the $\kappa$--symmetry 
projector \p{17} is similar to the r.h.s. of the M--theory superalgebra \p{19}.
This similarity is not accidental. It reflects the fact that $\kappa$--symmetric
5--brane configurations preserve half the supersymmetry of the $D=11$ vacuum and 
prompts a general form of the corresponding M--theory superalgebra projector.
The structure of this projector is easy to understand with the example of a 
planar five--brane stretched along first five coordinates of $D=11$ space--time
$y^i=X^i~(i=1,...,5)$, $y^0=X^0=a(y)$ and having a constant gauge field strength
$H_{ijk}$. Then the two--form and the five--form central charge have the 
following densities per spatial unit volume of the 5--brane
\begin{equation}\label{27}
Z^{i_1i_2}=-H^{*i_1i_2}, \qquad Y^{i_1...i_5}=\varepsilon^{i_1...i_5}.
\end{equation}
 From the similarity with the $\kappa$--symmetry projector \p{17} we 
derive that this 5--brane configuration will preserve 1/2 supersymmetry 
if its energy density is equal to the DBI Lagrangian and it has a 
non--zero spatial momentum
\begin{equation}\label{28}
P^0=\sqrt{\det(\d_{ij}+H^*_{ij})}, \qquad P^i={1\over 
8}\varepsilon^{ij_1j_2j_3j_4}H^*_{j_1j_2}H^*_{j_3j_4}.
\end{equation}
Note that these are exactly the values of the components of the 
canonical momentum of the planar 5--brane derived from the variation of 
the 5--brane Lagrangian with respect to the 5--brane velocity 
$\partial_0X^{\u m}$
$$
P_{\u m}={1\over{\sqrt{-g_6}}}{{\delta L}\over{\delta(\partial_0X^{\u 
m})}}.
$$
We see that this BPS 5--brane configuration is not static and moves 
along a direction orthogonal to $H^{*ij}$. From the point of view of 
intersecting branes this single 5--brane configuration with non--zero 
chiral field can be interpreted as a system of a purely magnetic 
5--brane (with $B^{(2)}=0$) and of two intersecting membranes moving
inside the 5--brane in a direction orthogonal to both of them.

Let us now turn to the consideration of 5--brane equations of motion which
follow from the action \p{14} and compare  
them \cite{pst3} with 5--brane equations
which arise in a superembedding approach to the description of 
on--shell worldvolume dynamics of the 5--brane \cite{hsw}.

The action \p{14} yields the following equations for the fields $B_{mn}(y)$,
$X^{\ m}(y)$ and $\Theta^\a(y)$, respectively:
\begin{equation}\label{B}
\hat H_{mnp}v^p=
{{\hat H^*_{mn}-{1\over 2}tr(\hat H^*)^2\hat H^*_{mn}+
(\hat H^*)^3_{mn}}\over{L_{DBI}}},
\end{equation} 
$$
{1\over 2}T^{mn}D_mE_n{}^{\underline{a}}=
$$
\begin{equation}\label{X}
={\varepsilon^{m_1\cdots m_6}\over\sqrt{-g}}\left[{1\over 6!}
F^{\underline{a}}{}_{m_6\cdots m_1} -{1\over (3!)^2}\left( 
F^{\underline{a}}{}_{m_6 m_5 m_4}\hat H_{m_3 m_2 m_1}
-E_{n}^{\underline{a}}
F^{n}{}_{m_6 m_5 m_4}\hat H_{m_3 m_2 m_1}\right)\right], 
\end{equation}
\begin{equation}\label{Th}
E^\a_mJ^m_{\a\b}=0=E_mJ^m(1-\bar \G).
\end{equation}
where
\begin{equation}\label{T}
T^{mn}={-4\over \sqrt{-g}}{\delta
S\over \delta g_{mn}}= 2 g^{mn} \left({{L}_{DBI}} 
-{1\over 4}\hat H^*_{mn}\hat H^{mn} \right)
- {1\over 2}\hat H^{mpq}\hat H^{*n}_{~~pq}
\end{equation}
is an ``energy--momentum'' tensor of the 5--brane,
and
\begin{equation}\label{J}
J^m=T^{mn}\G^n+2\G^m\G^{(6)}-H^{*mnp}\G_{np}, \quad J^m(1+\bar\G)\equiv 0,
\quad \G^{(6)}={1 \over 6! \sqrt{-g_6} }
\varepsilon^{m_1 ... m_6}\Gamma_{m_1... m_6}.
\end{equation}
Remember that $E^{\u a}_m(X,\Theta)$ and $E^\a_m(X,\Theta)$ 
are components of a
$D=11$ supervielbein pulled back into the 5--brane worldvolume,
and in a flat D=11 superspace 
$E^{\u a}_m=\partial_mX^{\u a}+i\partial_m\bar\Theta\Gamma^{\u a}\Theta$
and $E^\a_m=\partial_m\Theta^\a$.

The annihilation of $J^m$ by $1+\bar\G$ and, hence, the presence of the
projector $1-\bar \G$ in the $\Theta$--equaiton \p{Th}
is a consequence of $\kappa$--symmetry.
An alternative form of the 5--brane equations of motion was derived
in \cite{hsw} not from an action but 
by applying the superembedding approach \cite{bpstv}
to the description of the 5--brane as a supersurface embedded into $D=11$
curved target superspace. The embedding is specified by a geometrical 
condition which contains dynamical 5--brane equations as its consequences.
The $B^{(2)}$ - field equation, or a self--duality condition, is:
\begin{equation}\label{B1}
\hat H_{mnp}=3\partial_{[m}B_{np]}-A_{mnp}=4(m^{-1})_m^{~l}h_{lnp},
\end{equation}
$$
\hat H^*_{mnp}={4\over{(1-{2\over 3}tr{k^2})}}m_m^{~l}h_{lnp},
$$
where $h_{mnp}=h^*_{mnp}$ is an auxiliary self--dual tensor field,
which is present in the covariant superembedding approach instead
of the scalar field $a(y)$ (or $v_m(y)$) of the action formulation, and
\begin{equation}\label{m}
m_{mn}=g_{mn}+2h_m^{~pq}h_{npq}\equiv g_{mn}+2k_{mn}.
\end{equation}  
The $X^{\u m}(y)$--field equation has the following form
$$
{-{(m^2)^{mn}}\over{1-{2\over 3}tr{k^2}}}D_mE_n{}^{\underline{a}}=
$$
\begin{equation}\label{X1}
={\varepsilon^{m_1\cdots m_6}\over\sqrt{-g}}
\left[{1\over 6!}
F^{\underline{a}}{}_{m_6\cdots m_1} -{1\over (3!)^2}\left( 
F^{\underline{a}}{}_{m_6 m_5 m_4}\hat H_{m_3 m_2 m_1}
-E_{n}^{\underline{a}}
F^{n}{}_{m_6 m_5 m_4}\hat H_{m_3 m_2 m_1}\right)\right],
\end{equation}
and the $\Theta^\a(y)$ equation is
\begin{equation}\label{Th1}
E^\a_m\hat J^m_{\a\b}=0,
\end{equation}
where
\begin{equation}\label{J1}
\hat J^m=(1-\hat\G^T)\G^nm_n^{~m}(1-{1\over 6}h^{npq}\G_{npq}), 
\qquad \hat J^m(1+\hat \G)\equiv 0,
\end{equation}
\begin{equation}\label{hg}
(\hat\G_{\a\b})^T=\hat\G_{\b\a},\qquad 
\hat\G=\G^{(6)}+{1\over 3}h^{npq}\G_{npq}, \qquad (\hat\G)^2=1.
\end{equation}

The matrix $1+\hat\G$ plays the role of a $\kappa$-symmetry projector 
analogous to that of $1+\bar\G$ (eqs. \p{16}, \p{17}) in the action 
formulation. Note that eq. \p{Th1} is invariant under the following
transformations
\begin{equation}\label{k1}
\delta_{\hat\kappa}Z^ME_M^{\a}=(1+\hat\G)^\a_{~\b}\hat\k^\b
=\hat\k^\b(1+\hat\G^T)^{~\a}_{\b}.
\end{equation}

To establish the relationship between the equations \p{B}--\p{J}
and \p{B1}--\p{J1} \cite{pst3} one should, for example, eliminate $h_{mnp}$
from \p{B1}--\p{J1} and replace it with $H^*_{mn}\equiv H^*_{mnp}v^p$,
or vice versa. This can be done by use of eqs. \p{B1} and
the identity
\begin{equation}\label{id}
X^{mnl}=-{1\over2}{\epsilon^{mnlpqr}\over \sqrt{-g_6}}v_p X^*_{qrs}v^s
-3 v^{[m}X^{~nl]}_pv^p
\end{equation} 
which is valid for any $d=6$ three--form field.
Then it can be shown \cite{hsw1} that upon algebraic manipulations
eqs. \p{B1}, \p{id} produce the self--duality condition \p{B}.
The similarity of the $X$--field equations implies that
\begin{equation}\label{TT}
T^{mn}=-{{2(m^2)^{mn}}\over{1-{2\over 3}tr{k^2}}},
\end{equation}
which can be checked directly.
To relate the $\Theta$--equations \p{Th} and \p{Th1} one should notice that
the two projectors satisfy the identity
$$
{1 + \bar\Gamma
\over 2}~~{1 + \hat\Gamma \over 2} = {1 + \hat\Gamma
\over 2}.
$$
This implies that $J^m$ of \p{Th} and \p{J} is annihilated by both these
projectors
$$
J^m(1 + \bar\Gamma)=J^m(1 + \hat\Gamma)=0,
$$ 
and, hence, transformations \p{k1} can play the role of 
$\kappa$-symmetry transformations 
instead of \p{16}. This demonstrates that the $\kappa$--symmetry projector 
is not uniquely defined.

To complete the relationship of the $\Theta$--equations \p{Th} and \p{Th1}
one should just check that
$$
J^m(1 - \bar\Gamma)=J^m(1 - \hat\Gamma)=-{4\over{1-{2\over 3}tr{k^2}}}
\hat J^m,
$$
 from which it follows that eqs. \p{Th} and \p{Th1} coincide up to a
nonvanishing scalar factor.

We have now demonstrated that the 5--brane equations of motion
obtained from {\it a priori} different approachs are the same, which
testifies to the fact that they indeed describe classical dynamics of 
one and the same extended object. In this respect it 
would be of interest to understand more profound relationship between
the two 5--brane formulations.

We have also seen that the Noether supercharges derived from the 
5--brane action generate the M--theory superalgebra with both the 2--form
and the 5--form central charge, and that the knowledge of the action
allows one to obtain the explicit form of these topological charges in
terms of 5--brane coordinates and the worldvolume gauge field.

As has been shown in \cite{alwis,bbs} the 5--brane action admits a nonlocal
coupling to a $D=11$ supergravity action, which might be useful for studying
anomalies of M-theory in the presence of M--branes \cite{alwis,w,ano}.

Finally, recently  
the 5--brane action was used to construct 
a new interacting d=6 conformal field theory described
by a gauge--fixed worldvolume action of a 5--brane propagating in a $D=11$ 
baground of anti-de-Sitter geometry \cite{k}. This fits into a picture where
branes appear as boundaries of anti--de--Sitter superspaces and produce
$p+1$--dimensional superconformal field theories dual in a certain sense 
to $p+2$-dimensional adS supergravities.

An interesting and important problem is the quantization of the five--brane
and, in general, the quantization of the self--dual fields \cite{w}.
Here the problem of covariance appears once again.
In the approach considered above it is caused by topological 
piculiarities of the field $a(x)$, and by the necessity to gauge fix 
$a(x)$--field symmetry \p{5}, which in general breaks manifest 
space--time covariance of the action, as the temporal or the spatial 
gauge does. Therefore, after quantization one should check once again 
that quantum theory is space--time invariant. Recently an $SL(6,Z)$ 
modular invariant partition function for the M5--brane compactified on a 
six--torus was computed in \cite{dn}.

\bigskip
\bigskip
\noindent
{\bf Acknowledgements}. 
Results reviewed in this talk were obtained in a fruitful collaboration
with I. Bandos, E. Bergshoeff, N. Berkovits, K. Lechner, A. Nurmagambetov,
P. Pasti, M. Tonin and P. K. Townsend.
The author is grateful to the Organizers
of the Conference on Superfivebranes and Physics in 5+1 Dimensions (Trieste,
April 1--3, 1998), and of the INTAS Meeting (Paris, May 1--3, 1998) 
for hospitality and the A. von Humboldt Foundation 
for a financial support. I would also like to thank Kurt Lechner, Paolo Pasti
and Mario Tonin for valuable discussions and kind hospitality at Padua
University where this work was completed.

\end{document}